\documentclass[prd,twocolumn,showpacs,floatfix,superscriptaddress]{revtex4}

\usepackage{graphicx}
\usepackage{amsmath}
\usepackage{amssymb}
\usepackage{bm}
\begin{document}

\title{Bulk viscosity of strange quark matter: Urca versus non-leptonic processes}

\author{Basil A. Sa'd}
\email{sad@fias.uni-frankfurt.de}
\affiliation{%
Frankfurt Institute for Advanced Studies,
J.W. Goethe-Universit\"{a}t,
D-60438 Frankfurt am Main, Germany}%

\author{Igor A. Shovkovy}
\email{i-shovkovy@wiu.edu}
\altaffiliation[on leave
       of absence from ]{%
       Bogolyubov Institute for Theoretical Physics,
       03143, Kiev, Ukraine}%
\affiliation{%
Department of Physics, 
Western Illinois University, 
Macomb, IL 61455, USA}%

\author{Dirk H.\ Rischke}
\email{drischke@th.physik.uni-frankfurt.de}
\affiliation{%
Frankfurt Institute for Advanced Studies,
J.W. Goethe-Universit\"{a}t,
D-60438 Frankfurt am Main, Germany}%

\affiliation{Institut f\"ur Theoretische Physik,
J.W.\ Goethe-Universit\"at,
D-60438 Frankfurt am Main, Germany}

\date{\today}

\begin{abstract}
A general formalism for calculating the bulk viscosity of strange quark matter 
is developed. Contrary to the common belief that the non-leptonic processes
alone give the dominant contribution to the bulk viscosity, the inclusion of the 
{\rm Urca} processes is shown to play an important role 
at intermediate densities when the characteristic
r-mode oscillation frequencies are not too high. 
The interplay of 
non-leptonic and {\rm Urca} processes is analyzed in detail.
\end{abstract}



\maketitle

\section{Introduction}
\label{sec0}

Compact (neutron) stars provide a natural laboratory of matter under 
extreme conditions. In the central regions of such stars the baryon 
density of matter could reach values up to 10 times the nuclear 
saturation density (i.e., $10 \rho_0$ where $\rho_0\simeq 
0.15~\mbox{fm}^{-3}$). At such high density matter is likely to 
be in a deconfined state in which quarks rather than hadrons are the 
natural dynamical degrees of freedom \cite{Ivanenko1965,Ivanenko1969,
Itoh1970,Iachello1974,Collins1975}. 

One can argue that the ground state of deconfined quark matter is a color 
superconductor. (For reviews on color superconductivity, see Refs.~\cite{RajWil,
Alfordreview,Reddy2002,Rischke2003,Buballa2003,Huangreview,Shovkovy2004,
Alford:2006fw}). Many phases of color superconductivity are known 
that could possibly be realized in dense matter. It remains unclear, however, 
which of these describe the ground state of matter under the specific conditions 
in stars. This is because of theoretical uncertainties in treating the 
strongly coupled, non-perturbative dynamics in QCD at the baryon densities of relevance.

In order to clarify this one could use 
observational data from stars to narrow down 
the range of possibilities. For this program to work, one requires a 
detailed knowledge of the physical properties of various phases 
of matter that are likely to exist in stars. The transport properties
are most suitable for developing sufficiently sensitive, as well as 
unique and verifiable predictions regarding observational signals from 
stars. In this paper, we study the bulk viscosity of the normal phase 
of three-flavor quark matter. This work extends our recent analysis 
of the bulk viscosity in two-flavor quark matter
\cite{SSR1} by including strange quarks. This
 also augments existing studies of the topic \cite{Wang1984,
Sawyer2,Madsen,Xiaoping:2005js,Xiaoping:2004wc,Zheng:2002jq,
Dai:1996fe,Alford:2006gy}, in which the role of the Urca processes 
was not thoroughly investigated. 

\begin{figure}
\noindent
\includegraphics[width=0.4\textwidth]{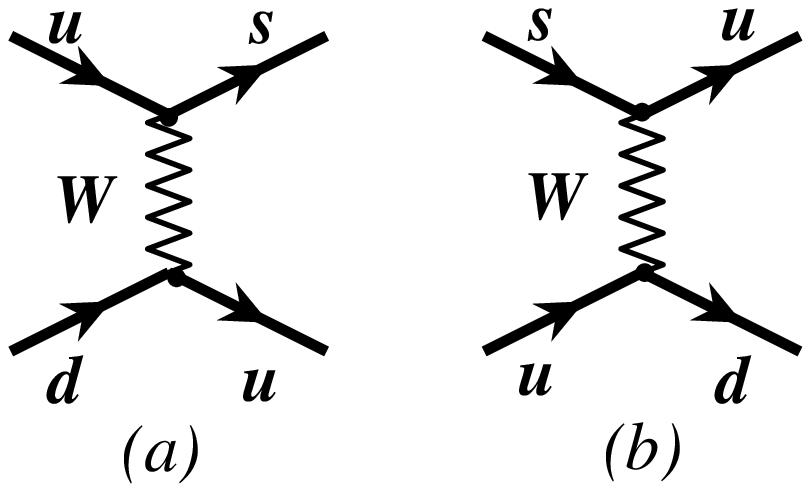}\\
\includegraphics[width=0.4\textwidth]{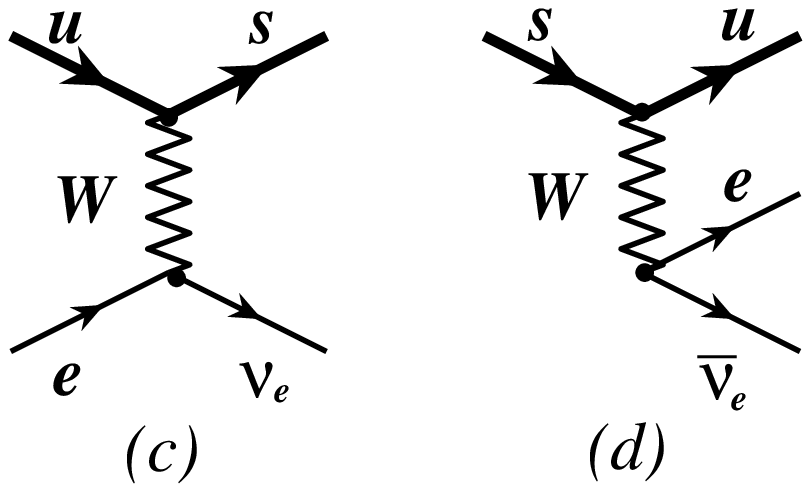}\\
\includegraphics[width=0.4\textwidth]{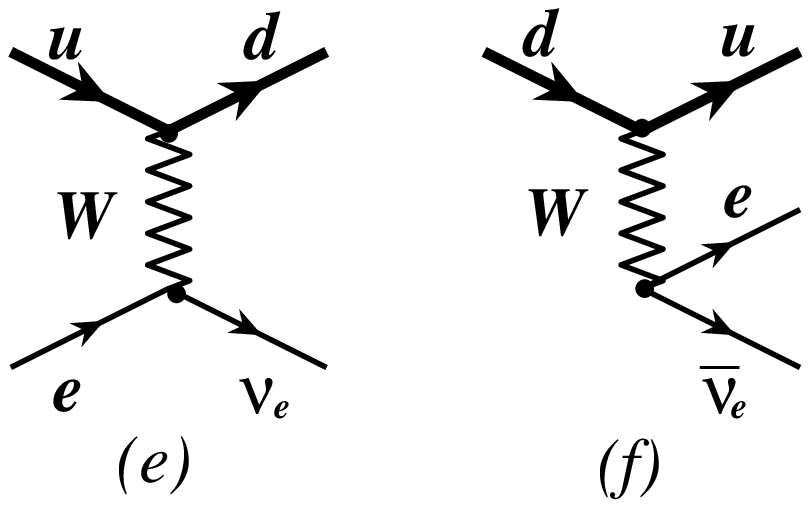}
\caption{Diagrammatic representation of the $\beta$ processes
that contribute to the bulk viscosity of quark matter in stellar cores.}
\label{fig-Urca_d_u_e}
\end{figure}

In general, the bulk viscosity is a measure of the kinetic energy dissipation 
during expansion and compression of a fluid. In compact stars, the
density oscillations of interest have characteristic frequencies that 
are of the same order of magnitude as the stellar rotation frequencies.
These are bound from below and from above, $1~\mbox{s}^{-1} 
\lesssim \omega\lesssim 
10^{3}~\mbox{s}^{-1}$. (For the fastest-spinning pulsar currently known, 
PSR J1748-2446ad, one has $\omega\approx 4.5\times 10^{3}~\mbox{s}^{-1}$ 
corresponding to $\nu=716~\mbox{Hz}$ \cite{716Hz}.)  The most important 
microscopic processes that provide the energy dissipation on the 
corresponding time scales are weak processes. Under conditions 
in stars, in particular, the 
bulk viscosity of quark matter is determined by the combined effect 
of the flavor-changing weak processes diagrammatically shown in 
Fig.~\ref{fig-Urca_d_u_e}. When an instantaneous departure from 
chemical equilibrium is induced by expansion/compression of 
matter, the weak processes try to restore the equilibrium state and, 
while doing this, reduce the oscillation energy. 

It is commonly argued that the bulk viscosity in the normal phase of 
three-flavor quark matter is dominated by the non-leptonic weak 
processes $u + d \leftrightarrow u + s$ \cite{Wang1984,Sawyer2,Madsen,
Xiaoping:2005js,Xiaoping:2004wc,Zheng:2002jq,Dai:1996fe,Alford:2006gy}.
These are shown diagrammatically in Figs.~\ref{fig-Urca_d_u_e}(a) and 
\ref{fig-Urca_d_u_e}(b). As for the Urca (semi-leptonic) processes, see
Figs.~\ref{fig-Urca_d_u_e}(c)--\ref{fig-Urca_d_u_e}(f), they have 
considerably lower rates which are suppressed parametrically by a 
factor of order $(T/\mu_e)^2$. This is in contrast to 
two-flavor (non-strange) quark matter, in which case the Urca 
processes $u + e^{-} \to d + \nu$ and $d \to u + e^{-}+\bar\nu$ 
[see Figs.~\ref{fig-Urca_d_u_e}(e) and \ref{fig-Urca_d_u_e}(f)] 
are the only ones that  contribute \cite{SSR1}. (For studies 
of the viscosity in various phases of dense nuclear matter, see 
Refs.~\cite{FlowersItoh1,FlowersItoh2,Sawyer,Jones1,Lindblom1,
Lindblom2,Drago1,Haensel1,Haensel2,Chat}.) 

In this paper, we study the bulk viscosity of three-flavor (strange) 
quark matter. One of the key issues that we address here is the interplay 
between the Urca and the non-leptonic processes. We shall show that, 
because a resonance-type phenomenon determines the bulk 
viscosity and because there is a subtle interference of the two weak processes, 
the simple argument about the dominance of the non-leptonic processes 
is not always justified.

This paper is organised as follows. In the next section, we develop the 
general formalism for calculating the bulk viscosity in three-flavor quark 
matter, paying special attention to the interplay of several different weak 
processes. In Sec.~\ref{bulk-vis-normal} we calculate the bulk viscosity 
in the normal phase of strange quark matter. In the same section, we also 
study the role of the {\rm Urca} processes on the bulk viscosity. The 
discussion of the results is given in Sec.~\ref{Conclusion}.

\section{Bulk viscosity in strange quark matter}
\label{bulk-viscosity}

In this section, we give a general derivation of the bulk viscosity resulting 
from the combined effect of all weak processes shown diagrammatically 
in Fig.~\ref{fig-Urca_d_u_e}. Following the same method as in Ref.~\cite{SSR1}, 
one can relate the bulk viscosity $\zeta$ to the average energy-density dissipation 
in pulsating matter,
\begin{equation}
\langle{{\cal \dot{E}}_{\rm diss}}\rangle 
=-\frac{\zeta \omega^2}{2}\left(\frac {\delta n_0}{n}\right)^{2}.
\label{zeta-def}
\end{equation}
In the derivation of this relation, we assumed that small oscillations
of the quark matter density are described by $\delta n = \delta n_{0}\, 
\mbox{Re}(e^{i\omega t})$ where $\delta n_{0}$ and $\omega$ are 
the magnitude and the frequency of the oscillations, respectively. 
The energy dissipation can be also expressed in terms of the induced 
pressure oscillations,
\begin{equation}
\langle \dot{\cal E}_{\rm diss}\rangle = \frac{n}{\tau} \int_0^{\tau} P \dot{V} dt
\label{diss-energy}
\end{equation}
where $V\equiv 1/n$ is the specific volume, and $\tau\equiv 2 \pi/\omega$ 
is the period of the oscillation. 

The density oscillations drive strange quark matter out of $\beta$ 
equilibrium. While the weak processes tend to restore 
equilibrium, they always lag behind. Thermal equilibrium, in contrast, 
is restored almost without any delay by very fast strong processes. 
The corresponding instantaneous quasi-equilibrium state is 
unambiguously characterised by the total baryon density $n$, 
the lepton fraction $X_e$,  and the strangeness fraction $X_s$,
\begin{subequations}
\begin{eqnarray}
n &=& \frac13 \left(n_u+n_d+n_s\right), \\
X_e &=& \frac{n_e}{n},\\
X_s &=& \frac{n_s}{n},
\end{eqnarray}
\label{composiiton}
\end{subequations}
where $n_u$, $n_d$, and $n_s$ are the number densities of up, down, 
and strange quarks, while $n_e$ is the number density of electrons. 
Because of the charge neutrality constraint, the number densities 
satisfy the following relation:
\begin{equation}
\frac23 n_u-\frac13 n_d-\frac13 n_s-n_e=0. 
\label{neutral}
\end{equation}
(Note that there are no microscopic processes that would lead to 
deviations from this charge neutrality constraint.) Using this 
constraint together with the definitions in Eq.~(\ref{composiiton}),
one can express the separate densities and, in fact, any thermodynamic 
quantity of quasi-equilibrium quark matter in terms of $n$, $X_e$, and 
$X_s$. For the number densities, for example, one finds
\begin{subequations}
\begin{eqnarray}
n_e &=& X_e n,   \label{rho_e}\\
n_u &=&\left(1+X_e\right) n,   \label{rho_u}\\
n_d &=&\left(2-X_e-X_s\right) n.  \label{rho_d}\\
n_s &=&X_s n.  \label{rho_s}
\end{eqnarray}
\end{subequations}
The number densities can also be expressed in terms of the corresponding 
chemical potentials, $n_i=n_i(\mu_i)$. In $\beta$ equilibrium, the chemical 
potentials are related as follows: $\mu_s=\mu_d=\mu_u+\mu_e$. In pulsating 
matter, on the other hand, the instantaneous departure from equilibrium 
is described by the following two independent parameters:
\begin{subequations}
\begin{eqnarray}
\delta\mu_{1} &\equiv& \mu_s-\mu_d =\delta\mu_s-\delta\mu_d ,\\
\delta\mu_{2} &\equiv& \mu_s-\mu_u-\mu_e =\delta \mu_s-\delta \mu_u-\delta \mu_e,
\label{delta-mu-def}
\end{eqnarray}
\end{subequations}
where $\delta\mu_{i}$ denotes the deviation of the chemical potential $\mu_i$
from its value in $\beta$ equilibrium. Note that the third possible combination 
of the chemical potentials, i.e., 
$\delta\mu_{3} \equiv \mu_d-\mu_u-\mu_e =\delta\mu_{2}-\delta\mu_{1}$, 
is not independent. 

In quasi-equilibrium, both quantities $\delta\mu_{1}$ and $\delta\mu_{2}$ 
can be expressed in terms of the variations of the three independent 
variables, $\delta n$, $\delta X_e$, and $\delta X_s$, 
\begin{eqnarray}
\delta\mu_{i} &=& C_{i} \frac{\delta n}{n} + B_{i} \delta X_e + A_{i}\delta X_s \, ,
\quad {(i=1,2)}
\label{delta-mu}
\end{eqnarray}
where, as follows from the definition, the coefficient functions are 
given by 
\begin{subequations}
\begin{eqnarray}
A_{1} &=& n \left(\frac{\partial \mu_d}{\partial n_d} 
                 + \frac{\partial \mu_s}{\partial n_s}\right), 
\label{def-A1}\\
B_{1} &=& n \frac{\partial \mu_d}{\partial n_d} , 
\label{def-B1}\\
C_{1} &=& n_s \frac{\partial \mu_s}{\partial n_s}
     -n_d \frac{\partial \mu_d}{\partial n_d},   \label{def-C}\\
A_{2} &=& n \frac{\partial \mu_s}{\partial n_s} ,
\label{def-A2}\\
B_{2} &=& -n \left(\frac{\partial \mu_u}{\partial n_u}
                 + \frac{\partial \mu_e}{\partial n_e} \right),
\label{def-B2}\\
C_{2} &=& n_s \frac{\partial \mu_s}{\partial n_s}
     -n_u \frac{\partial \mu_u}{\partial n_u}
     -n_e \frac{\partial \mu_e}{\partial n_e}.
\label{def-C2}
\end{eqnarray}
\label{def-ABC}
\end{subequations}
When the quantity $\delta\mu_{i}$ is non-zero, the two weak processes 
in the $i$th row of Fig.~\ref{fig-Urca_d_u_e} have slightly different rates. 
To leading order in $\delta\mu_{i}$, we could write 
\begin{subequations}
\begin{eqnarray}
\Gamma_{(a)} - \Gamma_{(b)} &=& - \lambda_{1} \delta\mu_{1} ,
\label{ratediff1}\\
\Gamma_{(c)} - \Gamma_{(d)} &=& - \lambda_{2} \delta\mu_{2} ,
\label{ratediff2}\\
\Gamma_{(e)} - \Gamma_{(f)} &=& - \lambda_{3} \left(\delta\mu_{2}-\delta\mu_{1}\right) ,
\label{ratediff3}
\end{eqnarray}
\label{ratediff0}
\end{subequations}
(Note that our conventions are such that the quantities $\lambda_{i}$ are 
non-negative.) The net effect of having different rates for each pair of the 
weak processes in Fig.~\ref{fig-Urca_d_u_e} is a change of the electron 
and strangeness fractions in the system, i.e.,
\begin{subequations}
\begin{eqnarray}
n \frac{d (\delta X_e)}{dt} &=& \lambda_{2} \delta\mu_{2} 
   + \lambda_{3} \left(\delta\mu_{2}-\delta\mu_{1}\right),
\label{ratediff11}\\
n \frac{d (\delta X_s)}{dt} &=& -\lambda_{1} \delta\mu_{1} - \lambda_{2} \delta\mu_{2}.
\label{ratediff22}
\end{eqnarray}
\label{ratediff}
\end{subequations}
According to these equations, the instantaneous values of the electron and strangeness 
fractions, i.e., $X_e\equiv X_e^{(0)}+\delta X_e$ and $X_s\equiv X_s^{(0)}+\delta X_s$, 
tend to approach their equilibrium values, $X_e^{(0)}$ and $X_s^{(0)}$, respectively. 
For example, a deficit of electrons (indicated by either $\delta\mu_2>0$ or 
$\delta\mu_2-\delta\mu_1>0$, or both) causes their production, see Eq.~(\ref{ratediff11}). 
Similarly, a surplus of strange quarks (indicated by either $\delta\mu_1>0$ or 
$\delta\mu_2>0$, or both) is to be removed through the weak processes, see 
Eq.~(\ref{ratediff22}). Because of finite rates for the weak processes, however, the 
oscillations of $\delta X_e$ and $\delta X_s$ always lag behind the density oscillations. 
In order to see this explicitly, we substitute $\delta\mu_{i}$ from Eq.~(\ref{delta-mu}) 
into Eq.~(\ref{ratediff}), and get the following set of equation for $\delta X_e$ and 
$\delta X_s$:
\begin{widetext}
\begin{subequations}
\begin{eqnarray}
n \frac{d (\delta X_e)}{dt} &=& 
\left[\left(\lambda_{2} +\lambda_{3} \right) C_{2}-\lambda_{3} C_{1}\right]
\frac{\delta n}{n}  
+\left[\left(\lambda_{2} +\lambda_{3} \right) B_{2}-\lambda_{3} B_{1}\right]
\delta X_e  
+\left[\left(\lambda_{2} +\lambda_{3} \right) A_{2}-\lambda_{3} A_{1}\right]
\delta X_s , \label{eqX_e}\\
n \frac{d (\delta X_e)}{dt} &=& 
-\left[\lambda_{1} C_{1}+\lambda_{2} C_{2}\right]\frac{\delta n}{n}  
-\left[\lambda_{1} B_{1}+\lambda_{2} B_{2}\right]\delta X_e  
-\left[\lambda_{1} A_{1}+\lambda_{2} A_{2}\right]\delta X_s.
\label{eqX_s}
\end{eqnarray}
\label{eqX_eX_s}
\end{subequations}
\end{widetext}
The periodic solution to this equation can be found most easily by making 
use of complex variables. Thus, by defining
\begin{subequations}
\begin{eqnarray}
\delta X_e &=& \mbox{Re}\left(\delta X_{e,0}\, e^{i\omega t}\right) ,\\
\delta X_s &=& \mbox{Re}\left(\delta X_{s,0}\, e^{i\omega t}\right),
\end{eqnarray}
\end{subequations}
we derive the following results for the complex magnitudes:
\begin{subequations}
\begin{eqnarray}
\delta X_{e,0} &=& \frac{\delta n_0}{n}\frac{d_{1}+i d_{2}}{g_{1}+i g_{2}},
\label{deltaX_e0}\\
\delta X_{s,0} &=& \frac{\delta n_0}{n}\frac{f_{1}+i f_{2}}{g_{1}+i g_{2}},
\label{deltaX_s0}
\end{eqnarray}
\label{deltaX_es0}
\end{subequations}
where  
\begin{subequations}
\begin{eqnarray}
&&\hspace{-10mm} d_{1} = \left(\alpha_{1} + \alpha_{2} + \alpha_{3} \right)
\left(A_{1}C_{2} - A_{2}C_{1}\right) ,\\
&&\hspace{-10mm} d_{2} =  \alpha_{1} \alpha_{2}\left(C_{2}-C_{1} \right)
+\alpha_{1} \alpha_{3} C_{2},\\
&&\hspace{-10mm} f_{1} = \left(\alpha_{1} + \alpha_{2} + \alpha_{3} \right)
\left(C_{1} B_{2} - C_{2} B_{1}\right),\\
&&\hspace{-10mm} f_{2} =  -\alpha_{1}\alpha_{3} C_{2} - \alpha_{2}\alpha_{3} C_{1},\\
&&\hspace{-10mm} g_{1} = -\alpha_{1}\alpha_{2}\alpha_{3}
+ \left(\alpha_{1} + \alpha_{2} + \alpha_{3} \right)
\left(B_{1} A_{2}-A_{1} B_{2}\right),\\
&&\hspace{-10mm} g_{2} = \alpha_{1}\alpha_{2} \left(B_{1} - B_{2}\right)
+\alpha_{1}\alpha_{3}\left(A_{2}-B_{2}\right)
+\alpha_{2}\alpha_{3} A_{1}, 
\end{eqnarray}
\end{subequations}
with $\alpha_{i}\equiv n \omega/\lambda_{i}$ ($i=1,2$). 
The pressure oscillations can be given in terms of the instantaneous 
values of $\delta n$, $\delta X_{e}$, and $\delta X_{s}$,
\begin{equation}
\delta P = \frac{\partial P}{\partial n}\, \delta n
+n\left(C_{1}-C_{2}\right) \delta X_e 
+nC_{1} \delta X_s , \label{press}
\end{equation}
where the $C_{i}$ are the same as in Eq.~(\ref{def-ABC}). In the derivation we 
took into account that $n_i =\partial P/\partial \mu_i$ and that the total 
pressure is given by the sum of the partial contributions of the quarks 
and electrons, $P=\sum_{i}P_{i}(\mu_i)$.

After taking into account the relation (\ref{press}) together with the 
solution for $\delta X_{e,0}$ and $\delta X_{s,0}$, see 
Eq.~(\ref{deltaX_es0}), the expression (\ref{diss-energy}) 
becomes
\begin{equation}
\langle \dot{\cal E}_{\rm diss}\rangle = -\frac{\omega}{2} \delta n_0
\left[ \left(C_{1}-C_{2}\right)\mbox{Im}\left(\delta X_{e,0}\right)
+C_{1}\mbox{Im}\left(\delta X_{s,0}\right) \right].
\label{diss}
\end{equation}
This shows that the imaginary parts of $\delta X_{e,0}$ and $\delta X_{s,0}$
are directly related to the dissipation of the energy density. From the physical 
point of view, these imaginary parts are related to the phase shifts (i.e.,
lagging) in the oscillations of the electron and strangeness fractions with 
respect to the oscillations of the density of quark matter.

Now, by comparing Eq.~(\ref{diss}) with the definition in Eq.~(\ref{zeta-def}), 
we derive an explicit expression for the bulk viscosity,
\begin{equation}
\zeta = \frac{n^2}{\omega \delta n_0 } 
\left[ \left(C_{1}-C_{2}\right)\mbox{Im}\left(\delta X_{e,0}\right)
+C_{1}\mbox{Im}\left(\delta X_{s,0}\right) \right] .
\end{equation}
Finally, after making use of the expressions for $\delta X_{e,0}$ and $\delta X_{s,0}$, 
see Eq.~(\ref{deltaX_es0}), this result can be rewritten as follows:
\begin{equation}
\zeta=\zeta_{1}+\zeta_{2}+\zeta_{3},
\label{zeta-general}
\end{equation}
where
\begin{widetext}
\begin{subequations}
\begin{eqnarray}
&&\hspace{-10mm} \zeta_{1} = \frac{n}{\omega}\frac{\alpha_{2}\alpha_{3}}{g_{1}^2+g_{2}^2}
\left[\alpha_{1}\alpha_{2}\alpha_{3} C_{1}^2
+\left(\alpha_{1}+\alpha_{2}+\alpha_{3}\right)
\left(A_{1}C_{2}-A_{2}C_{1}\right)\left(A_{1}C_{2}-(A_{1}-B_{1})C_{1}\right)\right],
\label{zeta1} \\
&&\hspace{-10mm} \zeta_{2} = \frac{n}{\omega}\frac{\alpha_{1}\alpha_{3}}{g_{1}^2+g_{2}^2}
\left[
\alpha_{1}\alpha_{2}\alpha_{3} C_{2}^2
+\left(\alpha_{1}+\alpha_{2}+\alpha_{3}\right)
\left[\left(A_{2}-B_{2} \right) C_{1} -(A_{1}-B_{1})C_{2}\right]
\left[\left(A_{2}-B_{2} \right) C_{1} -A_{2} C_{2}\right]\right],
\label{zeta2} \\
&&\hspace{-10mm}\zeta_{3} = \frac{n}{\omega}\frac{\alpha_{1}\alpha_{2}}{g_{1}^2+g_{2}^2}
\left[
\alpha_{1}\alpha_{2}\alpha_{3} \left(C_{1}-C_{2}\right)^2
+\left(\alpha_{1}+\alpha_{2}+\alpha_{3}\right)
\left(B_{1}C_{2}-B_{2}C_{1}\right)
\left[ (A_{1}-A_{2})C_{2}-(B_{2}-B_{1}+A_{1}-A_{2})C_{1}\right]\right].
\label{zeta3}
\end{eqnarray}
\label{zeta10-30}
\end{subequations}
This can be simplified using the relation $A_{1}-A_{2}\equiv B_{1}$, 
which follows from the definitions in Eq.~(\ref{def-ABC}). Therefore, we obtain
\begin{subequations}
\begin{eqnarray}
&& \zeta_{1} = \frac{n}{\omega}\frac{\alpha_{2}\alpha_{3}}{g_{1}^2+g_{2}^2}
\left[\alpha_{1}\alpha_{2}\alpha_{3} C_{1}^2
+\left(\alpha_{1}+\alpha_{2}+\alpha_{3}\right)
\left(A_{1}C_{2}-A_{2}C_{1}\right)^2\right],
\label{zeta11} \\
&&\zeta_{2} = \frac{n}{\omega}\frac{\alpha_{1}\alpha_{3}}{g_{1}^2+g_{2}^2}
\left[
\alpha_{1}\alpha_{2}\alpha_{3} C_{2}^2
+\left(\alpha_{1}+\alpha_{2}+\alpha_{3}\right)
\left[\left(A_{2}-B_{2} \right) C_{1} -A_{2}C_{2}\right]^2\right],
\label{zeta22} \\
&&\zeta_{3} = \frac{n}{\omega}\frac{\alpha_{1}\alpha_{2}}{g_{1}^2+g_{2}^2}
\left[
\alpha_{1}\alpha_{2}\alpha_{3} \left(C_{1}-C_{2}\right)^2
+\left(\alpha_{1}+\alpha_{2}+\alpha_{3}\right)
\left(B_{1}C_{2}-B_{2}C_{1}\right)^2\right].
\label{zeta33}
\end{eqnarray}
\label{zeta1-3}
\end{subequations}
\end{widetext}
Let us emphasize that the $\zeta_{i}$, $i=1,2,3$, are {\bf not} directly associated 
with the separate contributions of the three types of weak processes 
in Fig.~\ref{fig-Urca_d_u_e}. Because of an interference of the weak 
processes, each of these contributions depends on all three rates $\lambda_{i}$,
$i=1,2,3$.  The separation occurs {\bf only} in the high-frequency limit, i.e.,
$\alpha_{i}\equiv n \omega/\lambda_{i}\to \infty$. Indeed, in this case
\begin{equation}
\zeta_{i}\simeq \frac{\lambda_{i}}{\omega^2}C_{i}^2, \quad 
(\mbox{no sum over $i$}),
\end{equation}
where $C_{3}\equiv C_2-C_1$. Note that the formal criterion for 
this separation reads $\omega\gg \omega_{\rm sep}$. An estimate for
$\omega_{\rm sep}$ is derived from the parametric dependence of the 
coefficient functions $A_i$ and $B_i$ on densities,
\begin{equation}
\omega_{\rm sep} \simeq \frac{\sqrt{\lambda_{1}\left(\lambda_{2}+\lambda_{3}\right)}}{n_e^{2/3}}.
\label{omega-sep}
\end{equation}

In order to understand the general features of the result in Eq.~(\ref{zeta1-3}), 
let us consider the interplay of the non-leptonic and semi-leptonic contributions 
to the bulk viscosity. In the limit of vanishing rates for the semi-leptonic 
processes, i.e., $\lambda_2,\lambda_3\to0$, the bulk viscosity reduces to the 
following well-known result for strange quark matter \cite{Sawyer2,Madsen}:
\begin{equation}
\zeta_{\rm non} \simeq \frac{\lambda_{1}C_{1}^2}
{\omega^2+\left(\lambda_{1}A_{1}/n\right)^2}.
\label{zetaX00}
\end{equation}
In this limit, there are only non-leptonic processes left, and they induce 
the dissipation of the oscillation energy. 

In order to better understand the interplay of different types of processes, 
it is instructive to consider also the limit of an infinitely large non-leptonic rate, 
$\lambda_1\to \infty$, keeping the semi-leptonic rates $\lambda_2$ and 
$\lambda_3$ finite. In this case, the expression for the bulk viscosity is 
given by
\begin{equation}
\zeta_{\rm lep} \simeq \frac{\left(\lambda_{2}+\lambda_{3}\right)
\left(C_{2}-C_{1} A_{2}/A_{1}\right)^2}
{\omega^2+\left[\left(\lambda_{2}+\lambda_{3}\right)
\left(B_{2}-B_{1} A_{2}/A_{1}\right)/n\right]^2}.
\label{zetaIXX}
\end{equation}
It may appear puzzling that the two seemingly equivalent limits, 
namely $\lambda_{1}\gg \lambda_{2},\lambda_{3}$, lead to such very 
different results. The problem can be resolved by noting that 
the result in Eq.~(\ref{zetaX00}) is reliable only if the following 
additional constraint is satisfied: $\left(\lambda_{2}+\lambda_{3}\right)\ll n^{4/3}\omega^2/\lambda_{1}$.
[In deriving this constraint, we assumed that the coefficient functions
$A_i$ and $B_i$ scale with the density as $n^{1/3}$, which will turn out 
to be a reasonable approximation, see Eq.~(\ref{C-B-normal}) below.]
In contrast, the result in Eq.~(\ref{zetaIXX}) is reliable only if
$\lambda_{1}\gg n^{4/3}\omega^2/\left(\lambda_{2}+\lambda_{3}\right)$.
It is convenient, therefore, to introduce the following characteristic 
frequency:
\begin{equation}
\omega_{0} =\frac{\sqrt{\lambda_{1}\left(\lambda_{2}+\lambda_{3}\right)}}{n^{2/3}},
\label{omega0}
\end{equation}
which separates the two qualitatively different regimes. The result in 
Eq.~(\ref{zetaX00}) is to be used when the frequency of the density pulsations 
is sufficiently high, $\omega\gg \omega_{0}$, so that only the fast 
non-leptonic processes have a chance to dampen the kinetic energy 
efficiently. The result in Eq.~(\ref{zetaIXX}), on the other hand, should be 
used at sufficiently low frequencies, $\omega\ll \omega_{0}$, when only the 
much slower semi-leptonic processes provide a substantial damping 
of the oscillations. At intermediate frequencies,  $\omega\simeq \omega_{0}$, 
neither Eq.~(\ref{zetaX00}) nor Eq.~(\ref{zetaIXX}) provides a good approximation 
to the bulk viscosity because of a strong interference of the non-leptonic and 
semi-leptonic processes. Below we study this interference in some more detail.

\section{Bulk viscosity in normal phase}
\label{bulk-vis-normal}

In order to calculate the bulk viscosity in the normal phase of three-flavor 
quark matter, we need to determine the corresponding thermodynamic coefficients 
$A_{i}$, $B_{i}$ and $C_{i}$ [see Eq.~(\ref{def-ABC})] and calculate the 
difference of the rates of the three pairs of weak processes in 
Fig.~\ref{fig-Urca_d_u_e}. 

Let us start from the derivation of the coefficients $A_{i}$, $B_{i}$ and 
$C_{i}$. For this purpose, we use of the following zero-temperature 
expressions for the number densities of quarks and electrons:
\begin{widetext}
\begin{subequations}
\begin{eqnarray}
n_{f} &=& \frac{\left(\mu_{f}^2-m_{f}^2\right)^{3/2}}{\pi^2}
-\frac{2\alpha_s}{\pi^3}\mu_{f}\left(\mu_{f}^2-m_{f}^2\right)
\left(1-\frac{3m_{f}^2}{\mu_{f}\sqrt{\mu_{f}^2-m_{f}^2}}
\ln\frac{\mu_{f}+\sqrt{\mu_{f}^2-m_{f}^2}}{m_{f}}\right)
, \quad \mbox{for}\quad f=u,d,s,\\
n_{e}&=& \frac{1}{3\pi^2}\mu_e^3.
\end{eqnarray}
\label{despersion}
\end{subequations}
\end{widetext}
Note that the expressions for quarks include the leading-order 
corrections due to strong interactions \cite{Freedman1976,Baluni1977,
Fraga2004,SchaferChapter}. By making use of these relations together 
with the definitions in Eq.~(\ref{def-ABC}), we derive the following 
approximate expressions for the coefficient functions:  
\begin{subequations}
\begin{eqnarray}
A_{1} &\simeq & \frac{2\pi^2 n}{3\mu_d^2} ,
\label{A1-qm}\\
B_{1} &\simeq & \frac{\pi^2 n}{3\mu_d^2},
\label{B1-qm}\\
C_{1} &\simeq & -\frac{m_s^2-m_d^2}{3\mu_d} 
-C^{\prime}_{s}+C^{\prime}_{d},
\label{C1-qm}\\
A_{2} &\simeq & \frac{\pi^2 n}{3\mu_d^2},
\label{A2-qm}\\
B_{2} &\simeq & -\frac{\pi^2 n}{3}
\left(\frac{3}{\mu_e^2}+\frac{1}{\mu_u^2}\right) ,
\label{B2-qm}\\
C_{2} &\simeq& -\frac{m_s^2}{3\mu_s}+\frac{m_u^2}{3\mu_u} 
-C^{\prime}_{s}+C^{\prime}_{u}.
\label{C2-qm}
\end{eqnarray}
\label{C-B-normal}
\end{subequations}
On the right-hand side of these expressions, we made use of the equilibrium 
relations satisfied by the chemical potentials, $\mu_s=\mu_d=\mu_u+\mu_e$, 
and neglected higher-order mass corrections in the expressions for $A_{i}$ 
and $B_{i}$. In the expressions for $C_{i}$, on the other hand, we kept the 
leading-order mass corrections, which also include the contributions linear 
in the strong coupling constant $\alpha_s$,
\begin{equation}
C^{\prime}_{f}=
- \frac{4\alpha_s m_f^2}{3\pi\mu_f} 
\left(\ln\frac{2\mu_f}{m_f}-\frac23\right).
\end{equation}
The reason for keeping the mass corrections is that the functions $C_{i}$ 
vanish in the massless limit. In that limit, all $\zeta_i$'s
are zero as appropriate for scale invariant theories \cite{Son:2005tj}.

In order to further proceed with the calculation of the bulk viscosity 
in the normal phase of three-flavor quark matter, we also need to know 
the rate difference of the three pairs of weak processes in 
Fig.~\ref{fig-Urca_d_u_e}. The rates of both the non-leptonic and 
semi-leptonic processes have been calculated in the literature, 
see Refs.~\cite{Wang1984,Sawyer2,Madsenrate,Anand} and 
Refs.~\cite{Iwamoto,Iwamoto2}, respectively. In the limit of three 
massless quarks, for example, the rates are
\begin{subequations}
\begin{eqnarray}
\lambda_{1} &\simeq& \frac{64}{5\pi^3} G_F^2
                     \cos^2 \theta_C \sin^2 \theta_C \mu_d^5 T^2 ,
\label{lambda1}\\
\lambda_{2} &\simeq& \frac{17}{40\pi} G_F^2
                     \sin^2 \theta_C \mu_s m_s^2 T^4 ,                 
\label{lambda2}\\
\lambda_{3} &\simeq& \frac{17}{15\pi^2} G_F^2
                     \cos^2 \theta_C \alpha _s \mu_d \mu_u \mu_e T^4 .
\label{lambda3}
\end{eqnarray}
\label{lambda123}
\end{subequations}
When the value of the strange quark mass is not small, the expressions for the 
rates are more complicated \cite{Madsenrate,Anand} and can only be calculated numerically. For the purposes of this paper it suffices to use the approximate 
expressions in Eq.~(\ref{lambda123}). These provide a reasonable approximation 
for the range of strange quark masses considered below.

Now, by making use of the results for the rates (\ref{lambda123})  as well as for 
the coefficient functions in Eq.~(\ref{C-B-normal}), we can straightforwardly 
calculate the bulk viscosity. In the calculation, we use the following two 
representative sets of model parameters:

\begin{table}[h]
\begin{tabular}{|l||l|}
\hline
set A & set B\\
\hline
\hline
$n = 5\rho_0$ & $n = 10\rho_0$\\
$m_s = 300$~MeV & $m_s = 140$~MeV\\
$\alpha_s = 0.2$ & $\alpha_s  = 0.1$\\
\hline
\end{tabular}
\end{table}

In both cases the light quark masses are $m_u =  5$~MeV and
$m_d =  9$~MeV.
Regarding the choice of parameters for sets A and B, several 
comments are in order. Set B is supposed to be characteristic for the
conditions in the inner core of a quark star. 
The strong coupling constant $\alpha_s$ is small due
to asymptotic freedom 
and the in-medium constituent strange quark mass $m_s$
assumes a value close to its current value on account 
of chiral symmetry restoration at large baryon densities.
Set A applies to intermediate densities,
such as occur in the outer core of a quark star.
Here, the strong coupling constant $\alpha_s$
and the in-medium constituent strange quark mass $m_s$
assume larger values. 

\begin{table*}[!]
\begin{tabular}{|c|c||c|c|c||c|c||c|c||c|c|}
\hline
$n/\rho_0$   & $m_s$ {[MeV]}   & $\mu_e$ {[MeV]}  & $\mu_u$ {[MeV]}  & $\mu_d=\mu_s$ {[MeV]}   
& $A_1$ {[MeV]} & $A_2$ {[MeV]} 
& $B_1$ {[MeV]} & $B_2$ {[MeV]} 
& $C_1$ {[MeV]} & $C_2$  {[MeV]} \\
\hline
$5$ &$300$ &  $39.139$  & $402.463$  & $441.602$ 
& $239.432$ & $127.937$ 
& $111.386$ & $-3.726\times 10^4$ 
& $-60.463$ & $-60.460$ \\
$10$ & $140$ &  $7.396$  & $495.275$  & $502.671$ 
& $324.118$ & $164.288$ 
& $160.268$ & $-2.080\times 10^6$ 
& $-10.692$ & $-10.709$ \\
\hline
\end{tabular}
\caption{Two sets of parameters used in the calculation of the bulk viscosity. }
\label{TableI}
\end{table*}

It can be shown that the electron chemical potential is driven by both
the strange quark mass and the leading-order corrections due to 
strong interactions, see Eq.~(\ref{despersion}). Without  
$\alpha_s$ corrections, the electron chemical potential would be 
$9.9$ and $50.5$~MeV for strange quark masses of $140$~MeV 
($n=10\rho_0$) and $300$~MeV ($n=5\rho_0$), respectively. 
The leading-order corrections due to strong interaction tend to 
reduce the values of $\mu_e$. A simple analysis shows that, in fact, 
the electron chemical potential could even formally change the sign 
when the value of $\alpha_s$ is sufficiently large (e.g., $\alpha_s
\gtrsim 0.5$). This would mean that strange quark matter requires 
the presence of the positrons rather than electrons to stay 
neutral. While not forbidden, such a possibility should be accepted 
with great caution. Indeed, the leading-order $\alpha_s$ corrections 
may be unreliable in this regime.

The general algorithm for calculating the bulk viscosity is as follows.
First, by assuming a fixed value of the baryon density of neutral quark 
matter $n$ (i.e., $n= 5 \rho_0\approx 0.75~\mbox{fm}^{-3}$ or 
$n= 10 \rho_0\approx 1.5~\mbox{fm}^{-3}$ for the two cases considered), 
we determine the chemical potentials of the quarks and electrons in 
$\beta$ equilibrium. For the two representative sets, 
the values of the chemical potentials as well as the 
coefficient functions $A_{i}$, $B_{i}$ and $C_{i}$ are given in 
Table~\ref{TableI}. These are used in the calculation of the rates 
of the weak processes. Putting everything together, the result for 
the bulk viscosity follows from Eq.~(\ref{zeta-general}).

\begin{figure}[t]
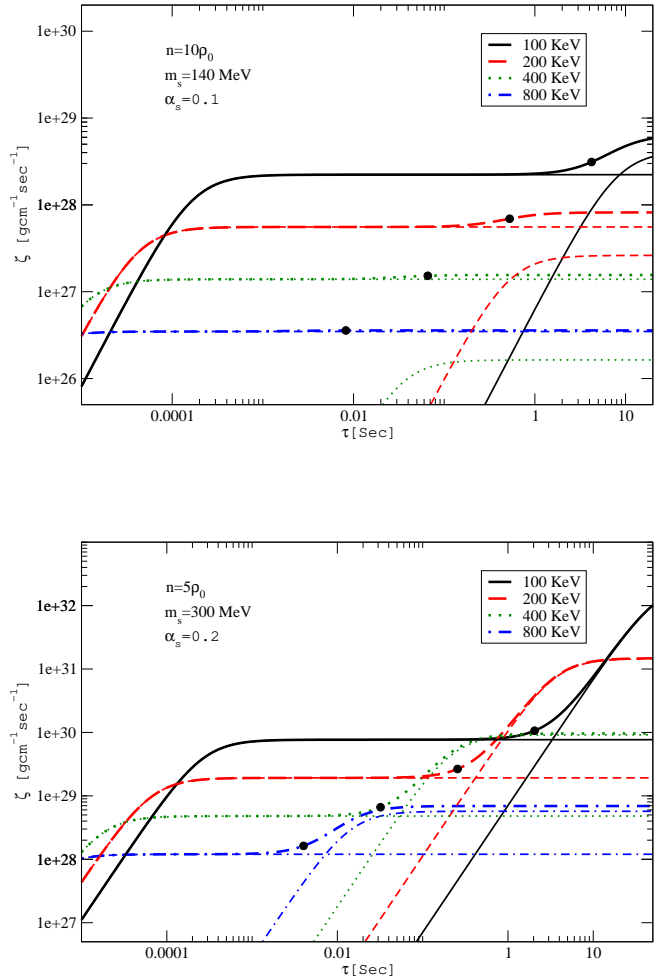

\includegraphics[width=0.48\textwidth]{zeta_tau_10_140_5_9_1.eps}\\
\vspace*{1.3 cm}
\includegraphics[width=0.48\textwidth]{zeta_tau_5_300_5_9_2.eps}
\vspace*{0.1 cm}
\caption{(Color online) The bulk viscosity for the normal phase of three-flavor 
quark matter as a function of the period of density oscillations.
Results for set B are shown in the upper panel and those for set
A in the lower panel. For each temperature, the dots 
on the lines correspond to the values of the frequency defined in 
Eq.~(\ref{omega0}).}
\label{bv-fig1}
\end{figure}

The numerical results for the bulk viscosity as a function of the period of 
density oscillations $\tau$ are presented in Fig.~\ref{bv-fig1} for the two 
cases: (i) $n= 10 \rho_0$ and $m_s=140$~MeV (upper panel), and  
(ii) $n= 5 \rho_0$ and $m_s=300$~MeV (lower panel). Different 
line types correspond to different values of the temperature:
$T=0.1$~MeV (solid lines), $T=0.2$~MeV (dashed lines),
$T=0.4$~MeV (dotted lines), and $T=0.8$~MeV (dashed-dotted lines).
The thin lines show the high- and low-frequency approximations, 
defined in Eqs.~(\ref{zetaX00}) and (\ref{zetaIXX}), for each value
of the temperature. 

Note the double-step structure of the bulk viscosity as a function of 
$\tau$. The first step at small $\tau$ is the 
usual one. It corresponds to the low-frequency saturation (with 
$\zeta_{\rm non}^{\rm (max)}\sim 1/\lambda_1$) of the non-leptonic 
contribution to the bulk vicosity, see Eq.~(\ref{zetaX00}). The second 
step at about $\tau=2\pi/\omega_0$ (marked by dots in the figures) is 
a qualitatively new feature. As should be clear from our discussion in the preceding section, 
its appearance is the consequence of the interplay between the non-leptonic 
and the semi-leptonic weak processes contributing to the bulk viscosity of 
strange quark matter. At sufficiently large $\tau$ (i.e., low frequencies) the 
contribution of the semi-leptonic processes also saturates.

The increase of the bulk viscosity at low frequencies due to the contributions 
of the slower semi-leptonic processes is not unexpected. (For the same 
reason, weak processes are much more important for compact stars 
than strong processes, operating on typical QCD time scales of order 
$1~\mbox{fm}/c$). However, the main observation here is that the increase of 
the bulk viscosity due to the interplay between the non-leptonic and semi-leptonic 
weak processes could already be visible at frequencies relevant for the physics 
of compact stars. Moreover, by comparing the results in the two panels of 
Fig.~\ref{bv-fig1}, we find that for
the conditions corresponding to lower densities, the range 
of frequencies where the semi-leptonic processes contribute 
widens significantly.

\begin{figure}[t]
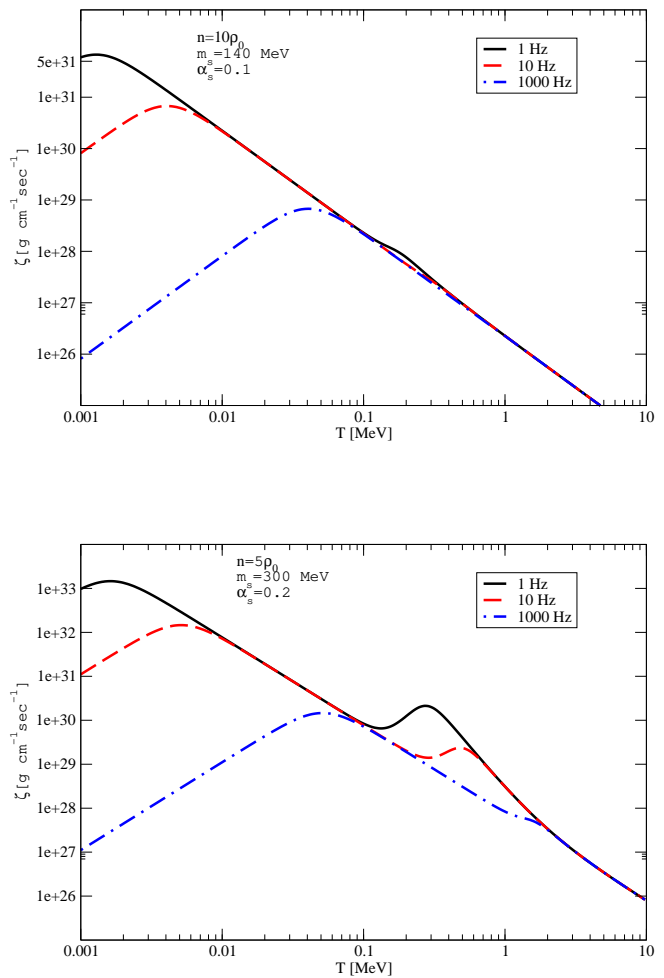

\vspace*{1.3 cm}
\includegraphics[width=0.48\textwidth]{zeta_T_10_140_5_9_1.eps}\\
\vspace*{1.3 cm}
\includegraphics[width=0.48\textwidth]{zeta_T_5_300_5_9_2.eps}
\vspace*{0.1 cm}
\caption{(Color online) The bulk viscosity for the normal phase of three-flavor 
quark matter as a function of the temperature for several fixed values
of the frequency of density oscillations. Results for set B
are shown in the upper panel and those for set A in the lower panel.}
\label{bv-fig2}
\end{figure}

The bulk viscosity as a function of the temperature is given in 
Fig.~\ref{bv-fig2} 
for set B (upper panel) and for set A (lower panel). 
The additional increase of the bulk viscosity due to the semi-leptonic processes
is seen as a ``bump" at intermediate values of the temperature. The results in 
the two panels demonstrate once again that the relative role of the semi-leptonic 
processes increases
with increasing period of density oscillations, provided the 
baryon density is sufficiently small (set A).

{From} Fig.~\ref{bv-fig2}, we find that the substantial modification of the bulk 
viscosity due to the {\rm Urca} processes occurs at temperatures between 
about $0.1$~MeV and $1$~MeV. This temperature range is of relevance to 
young neutron stars and, therefore, should be studied in more detail. 

\begin{figure}
\includegraphics[width=0.48\textwidth]{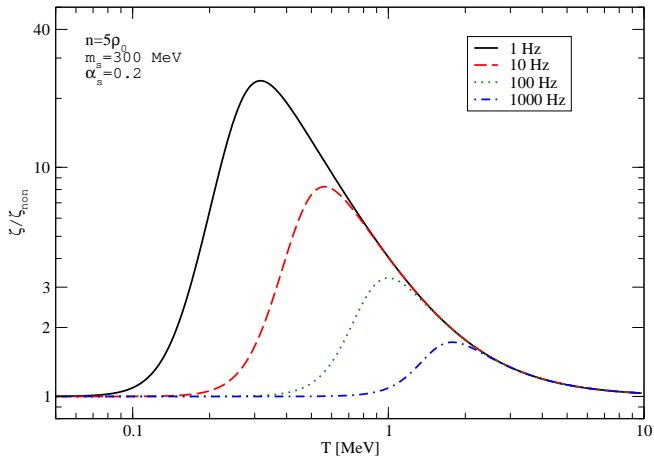}\\
\vspace*{0.2 cm}
\caption{(Color online) The ratio $\zeta/\zeta_{\rm non}$ as a function 
of temperature for set A. The results for three fixed 
values of the density oscillation frequencies are shown:
$\frac{1}{\tau}=1$~Hz (solid line), 
$\frac{1}{\tau}=10$~Hz (dashed line),  
$\frac{1}{\tau}=100$~Hz (dotted line), 
and $\frac{1}{\tau}=1000$~Hz (dashed-dotted line).}
\label{bv-fig3}
\end{figure}

Perhaps the best way to appreciate the relative role of the {\rm Urca} 
processes is to study the ratio between the complete expression for the 
bulk viscosity (\ref{zeta-general}) and the commonly used approximate 
form (\ref{zetaX00}) that takes only the non-leptonic interactions into account.
In the more interesting case of parameter set A, this ratio is shown in 
Fig.~\ref{bv-fig3} as a function of temperature for four different values 
of the period of density oscillations: 
$\frac{1}{\tau}=1$~Hz (solid line), 
$\frac{1}{\tau}=10$~Hz (dashed line),  
$\frac{1}{\tau}=100$~Hz (dotted line), 
$\frac{1}{\tau}=1000$~Hz (dashed-dotted line). 
The plot shows how the ratio $\zeta/\zeta_{\rm non}$ 
changes when the period of oscillations $\tau$ varies in the whole range 
from $1$~ms to $1$~s. 

The results presented in Fig.~\ref{bv-fig3} are the main results of this paper.
They show that neglecting the {\rm Urca} processes can result in 
underestimating the value of the bulk viscosity by an order of magnitude, 
at intermediate densities and sufficiently large in-medium strange quark
masses. 
We believe this finding 
might be of relevance for strange quark matter under conditions realized 
inside young neutron stars.

Before concluding this section, it is appropriate to note that the role of the 
semi-leptonic processes is negligible in the case of the smaller strange quark 
mass, $m_s=140$~MeV. This is seen from Fig.~\ref{bv-fig2} (upper panel), 
and this can also be confirmed by studying the ratio $\zeta/\zeta_{\rm non}$.
Its value deviates from 1 by at most $28\%$. This is in qualitative agreement 
with earlier studies \cite{Wang1984,Sawyer2,Madsen,Xiaoping:2005js,
Xiaoping:2004wc,Zheng:2002jq,Dai:1996fe,Alford:2006gy}, neglecting the 
semi-leptonic processes.

\section{Discussion}
\label{Conclusion}

In this paper we have studied the subtle interplay between the {\rm Urca} 
(semi-leptonic) and the non-leptonic weak processes in determining the 
bulk viscosity of neutral, $\beta$-equilibrated strange quark matter. In 
general, the contributions of the two types of weak processes 
are not separable. The exception is the high-frequency limit, $\omega 
\gg \omega_{\rm sep}$, see Eq.~(\ref{omega-sep}), in which case the 
contributions naturally separate. Because of a much higher rate of the 
non-leptonic processes, they dominate in this high-frequency limit. 

With decreasing frequency the role of the {\rm Urca} processes 
gradually increases. The maximal mixing with the non-leptonic processes 
occurs at about the frequency $\omega_0$, see Eq.~(\ref{omega0}) for
the definition. Depending on the specific choice of model parameters,
typical values of $\omega_0$ are in the range from about $1~\mbox{s}^{-1}$ 
to $10^{3}~\mbox{s}^{-1}$, see also Fig.~\ref{bv-fig1} for the corresponding 
values of the period of oscillations.

Our numerical results for the normal-conducting phase of strange quark matter 
demonstrate that the commonly used approximation, in which the {\rm Urca} 
processes are completely neglected, could substantially underestimate the 
value of the bulk viscosity, see Figs.~\ref{bv-fig1} and \ref{bv-fig2}. By 
sweeping over a wide range of model parameters, we find that the role of 
the {\rm Urca} processes is most important in the range of temperatures 
between about $0.1$~MeV and $1$~MeV. In the outer core of a 
quark star, i.e., when the density is not too large,
the strange quark mass not too small, and/or the period of 
oscillations not 
too short, the inclusion of the {\rm Urca} processes could lead to an increase 
of the viscosity by an order of magnitude, see Fig.~\ref{bv-fig3}. 

The findings of this work could have important implications for the physics
of young neutron stars with strange quark matter interiors and/or for 
pure strange stars that could potentially exist too. In connection to the 
r-modes instabilities driven by gravitational radiation, the increase of the 
bulk viscosity due to the {\rm Urca} processes is likely to broaden the region 
of stellar stability in the temperature--angular-frequency plane. The fact that 
the largest change occurs for the temperature range from about $0.1$~MeV 
to $1$~MeV might be also very important. Indeed, at the lower end of this 
range, the dominant role in suppressing the r-modes is expected to pass 
from the bulk viscosity to the shear viscosity. If the bulk viscosity is 
substantially higher than previously estimated, it may dominate the 
dissipative dynamics to much lower temperatures. 

In view of possible color superconductivity in strange quark matter, in the 
future one should also investigate the role of the {\rm Urca} processes in 
color-superconducting phases with various types of spin-zero and spin-one 
Cooper pairing. Because of the gaps in the quasiparticle spectra, the interplay 
between the semi- and non-leptonic processes in superconductors is expected 
to become more complicated due to the suppression of the rates. Certain general 
features of the corresponding dissipative dynamics could be predicted even 
without a detailed study. One could say, for example, that (i) the suppression of 
the rates should lead to the suppression of the bulk viscosity in the high-frequency 
limit, $\omega\gg \omega_{\rm sep}$, (ii) the border line itself, $\omega_{\rm sep}$, 
should shift to a lower value, (iii) the rates of the {\rm Urca} processes could 
possibly become even higher than the rates of the non-leptonic ones. The 
first two observations are simple consequences of our general results in 
Sec.~\ref{bulk-viscosity}. The situation described in item (iii) could be realized 
when all quark quasiparticles are gapped, for example, in one of the versions 
of the color-spin-locked phase. At low temperatures, the rates of the semi-
and non-leptonic processes are suppressed by exponential factors 
$\exp(-\phi/T)$ and $\exp(-2\phi/T)$, respectively. A more detailed discussion 
of spin-one color-superconducting phases of strange quark matter is subject of a subsequent paper \cite{SSR3}.

{\em Note added.} While writing our paper, we learned that another 
study of the bulk viscosity of strange quark matter is done by 
H.~Dong, N.~Su, and Q.~Wang \cite{bulkDSW}.

\section*{Acknowledgements}

The authors acknowledge discussions with M.~Alford, D.~Bandyopadhyay,
J.~Berges, M.~Buballa, F.P.~Esposito, E.~Ferrer, V.~Incera, J.~Noronha,
A.~Schmitt, J.~Wambach, and Q.~Wang. 
B.A.S. acknowledges support from the Frankfurt International Graduate School 
of Science (FIGSS). The work of D.H.R. and I.A.S. was supported in part by 
the Virtual Institute of the Helmholtz Association under grant No. VH-VI-041, 
by the Gesellschaft f\"{u}r Schwerionenforschung (GSI), and by the Deutsche 
Forschungsgemeinschaft (DFG).

\bibliography{bulk_vis_strange_v5}

\end{document}